 \documentclass[aps,prd,preprint,floatfix,nofootinbib,superscriptaddress,showpacs,amssymb]{revtex4}
 \usepackage{type1cm}
 \usepackage{graphicx}
\def\Z0{${\em Z^0\/}$}

\def\r#1 {$^{#1}$}

\newcommand{\gevc} { {\rm GeV/}c}
\newcommand{\gevcc}{ {\rm GeV/}c^2}

\def\gepsfcentered#1{
  \def\testit{#1}
  \def\lbracket{[}
  \ifx\testit\lbracket
    \let\dofilecmd=\gepsfwithopt
  \else
    \let\dofilecmd=\gepsfnoopt
  \fi
  \dofilecmd}

\def\gepsfnoopt#1{
  \begin{center}
  \leavevmode
  \epsffile{#1}
  \end{center}}

\def\gepsfwithopt#1 #2 #3 #4]#5{
  \begin{center}
  \leavevmode
  \gepsfmaxx=0.94\textwidth
  \epsffile[#1 #2 #3 #4]{#5}
  \end{center}}

\newdimen\gepsfmaxx
\def\epsfsize#1#2{
  \ifnum \epsfxsize=0
    \ifnum \epsfysize=0
      \ifnum #1 > \gepsfmaxx
        \gepsfmaxx
      \else
        #1
      \fi
    \else
      \epsfxsize
    \fi
  \else
    \epsfxsize
  \fi
}

\begin{document}
 \bibliographystyle{apsrev}

 %\hugehead
%\today
%\pagewiselinenumbers
%%%%%%%%%%%%%%%%%%%%%%%%%%%%%%%%%%%%%%%%%%%%%%%%%%%%%
 \title {Improved determination of the sample composition
 of dimuon events produced in {\boldmath $p\bar{p}$}
 collisions at {\boldmath $\sqrt{s}=1.96$} TeV 
}
%%%%%%%%%%%%%%%%%%%%%%%%%%%%%%%%%%%%%%%%%%%%%%%%%%%%%
 \affiliation{Institute of Physics, Academia Sinica, Taipei, Taiwan 11529, Republic of China} 
\affiliation{Argonne National Laboratory, Argonne, Illinois 60439, USA} 
\affiliation{University of Athens, 157 71 Athens, Greece} 
\affiliation{Institut de Fisica d'Altes Energies, ICREA, Universitat Autonoma de Barcelona, E-08193, Bellaterra (Barcelona), Spain} 
\affiliation{Baylor University, Waco, Texas 76798, USA} 
\affiliation{Istituto Nazionale di Fisica Nucleare Bologna, $^{aa}$University of Bologna, I-40127 Bologna, Italy} 
\affiliation{University of California, Davis, Davis, California 95616, USA} 
\affiliation{University of California, Los Angeles, Los Angeles, California 90024, USA} 
\affiliation{Instituto de Fisica de Cantabria, CSIC-University of Cantabria, 39005 Santander, Spain} 
\affiliation{Carnegie Mellon University, Pittsburgh, Pennsylvania 15213, USA} 
\affiliation{Enrico Fermi Institute, University of Chicago, Chicago, Illinois 60637, USA}
\affiliation{Comenius University, 842 48 Bratislava, Slovakia; Institute of Experimental Physics, 040 01 Kosice, Slovakia} 
\affiliation{Joint Institute for Nuclear Research, RU-141980 Dubna, Russia} 
\affiliation{Duke University, Durham, North Carolina 27708, USA} 
\affiliation{Fermi National Accelerator Laboratory, Batavia, Illinois 60510, USA} 
\affiliation{University of Florida, Gainesville, Florida 32611, USA} 
\affiliation{Laboratori Nazionali di Frascati, Istituto Nazionale di Fisica Nucleare, I-00044 Frascati, Italy} 
\affiliation{University of Geneva, CH-1211 Geneva 4, Switzerland} 
\affiliation{Glasgow University, Glasgow G12 8QQ, United Kingdom} 
\affiliation{Harvard University, Cambridge, Massachusetts 02138, USA} 
\affiliation{Division of High Energy Physics, Department of Physics, University of Helsinki and Helsinki Institute of Physics, FIN-00014, Helsinki, Finland} 
\affiliation{University of Illinois, Urbana, Illinois 61801, USA} 
\affiliation{The Johns Hopkins University, Baltimore, Maryland 21218, USA} 
\affiliation{Center for High Energy Physics: Kyungpook National University, Daegu 702-701, Korea; Seoul National University, Seoul 151-742, Korea; Sungkyunkwan University, Suwon 440-746, Korea; Korea Institute of Science and Technology Information, Daejeon 305-806, Korea; Chonnam National University, Gwangju 500-757, Korea; Chonbuk National University, Jeonju 561-756, Korea} 
\affiliation{Ernest Orlando Lawrence Berkeley National Laboratory, Berkeley, California 94720, USA} 
\affiliation{University of Liverpool, Liverpool L69 7ZE, United Kingdom} 
\affiliation{University College London, London WC1E 6BT, United Kingdom} 
\affiliation{Centro de Investigaciones Energeticas Medioambientales y Tecnologicas, E-28040 Madrid, Spain} 
\affiliation{Massachusetts Institute of Technology, Cambridge, Massachusetts 02139, USA} 
\affiliation{Institute of Particle Physics: McGill University, Montr\'{e}al, Qu\'{e}bec, Canada H3A~2T8; Simon Fraser University, Burnaby, British Columbia, Canada V5A~1S6; University of Toronto, Toronto, Ontario, Canada M5S~1A7; and TRIUMF, Vancouver, British Columbia, Canada V6T~2A3} 
\affiliation{University of Michigan, Ann Arbor, Michigan 48109, USA} 
\affiliation{Michigan State University, East Lansing, Michigan 48824, USA}
\affiliation{Institution for Theoretical and Experimental Physics, ITEP, Moscow 117259, Russia}
\affiliation{University of New Mexico, Albuquerque, New Mexico 87131, USA} 
\affiliation{Northwestern University, Evanston, Illinois 60208, USA} 
\affiliation{The Ohio State University, Columbus, Ohio 43210, USA} 
\affiliation{Okayama University, Okayama 700-8530, Japan} 
\affiliation{Osaka City University, Osaka 588, Japan} 
\affiliation{University of Oxford, Oxford OX1 3RH, United Kingdom} 
\affiliation{Istituto Nazionale di Fisica Nucleare, Sezione di Padova-Trento, $^{bb}$University of Padova, I-35131 Padova, Italy} 
\affiliation{LPNHE, Universite Pierre et Marie Curie/IN2P3-CNRS, UMR7585, Paris, F-75252 France} 
\affiliation{Istituto Nazionale di Fisica Nucleare Pisa, $^{cc}$University of Pisa, $^{dd}$University of Siena and $^{ee}$Scuola Normale Superiore, I-56127 Pisa, Italy} 
\affiliation{University of Pittsburgh, Pittsburgh, Pennsylvania 15260, USA} 
\affiliation{Purdue University, West Lafayette, Indiana 47907, USA} 
\affiliation{University of Rochester, Rochester, New York 14627, USA} 
\affiliation{The Rockefeller University, New York, New York 10065, USA} 
\affiliation{Istituto Nazionale di Fisica Nucleare, Sezione di Roma 1, $^{ff}$Sapienza Universit\`{a} di Roma, I-00185 Roma, Italy} 

\affiliation{Rutgers University, Piscataway, New Jersey 08855, USA} 
\affiliation{Texas A\&M University, College Station, Texas 77843, USA} 
\affiliation{Istituto Nazionale di Fisica Nucleare Trieste/Udine, I-34100 Trieste, $^{gg}$University of Udine, I-33100 Udine, Italy} 
\affiliation{University of Tsukuba, Tsukuba, Ibaraki 305, Japan} 
\affiliation{Tufts University, Medford, Massachusetts 02155, USA} 
\affiliation{University of Virginia, Charlottesville, Virginia 22906, USA}
\affiliation{Waseda University, Tokyo 169, Japan} 
\affiliation{Wayne State University, Detroit, Michigan 48201, USA} 
\affiliation{University of Wisconsin, Madison, Wisconsin 53706, USA} 
\affiliation{Yale University, New Haven, Connecticut 06520, USA} 
\author{T.~Aaltonen}
\affiliation{Division of High Energy Physics, Department of Physics, University of Helsinki and Helsinki Institute of Physics, FIN-00014, Helsinki, Finland}
\author{B.~\'{A}lvarez~Gonz\'{a}lez$^w$}
\affiliation{Instituto de Fisica de Cantabria, CSIC-University of Cantabria, 39005 Santander, Spain}
\author{S.~Amerio}
\affiliation{Istituto Nazionale di Fisica Nucleare, Sezione di Padova-Trento, $^{bb}$University of Padova, I-35131 Padova, Italy} 

\author{D.~Amidei}
\affiliation{University of Michigan, Ann Arbor, Michigan 48109, USA}
\author{A.~Anastassov}
\affiliation{Northwestern University, Evanston, Illinois 60208, USA}
\author{A.~Annovi}
\affiliation{Laboratori Nazionali di Frascati, Istituto Nazionale di Fisica Nucleare, I-00044 Frascati, Italy}
\author{J.~Antos}
\affiliation{Comenius University, 842 48 Bratislava, Slovakia; Institute of Experimental Physics, 040 01 Kosice, Slovakia}
\author{G.~Apollinari}
\affiliation{Fermi National Accelerator Laboratory, Batavia, Illinois 60510, USA}
\author{A.~Apresyan}
\affiliation{Purdue University, West Lafayette, Indiana 47907, USA}
\author{T.~Arisawa}
\affiliation{Waseda University, Tokyo 169, Japan}
% ...
\collaboration{CDF Collaboration\footnote{With visitors from $^a$University of MA Amherst,
Amherst, MA 01003, USA,
$^b$Istituto Nazionale di Fisica Nucleare, Sezione di Cagliari, 09042 Monserrato (Cagliari), Italy,
$^c$University of CA Irvine, Irvine, CA  92697, USA,
$^d$University of CA Santa Barbara, Santa Barbara, CA 93106, USA,
$^e$University of CA Santa Cruz, Santa Cruz, CA  95064, USA,
$^f$CERN,CH-1211 Geneva, Switzerland,
$^g$Cornell University, Ithaca, NY  14853, USA, 
$^h$University of Cyprus, Nicosia CY-1678, Cyprus, 
$^i$University College Dublin, Dublin 4, Ireland,
$^j$University of Fukui, Fukui City, Fukui Prefecture, Japan 910-0017,
$^k$Universidad Iberoamericana, Mexico D.F., Mexico,
$^l$Iowa State University, Ames, IA  50011, USA,
$^m$University of Iowa, Iowa City, IA  52242, USA,
$^n$Kinki University, Higashi-Osaka City, Japan 577-8502,
$^o$Kansas State University, Manhattan, KS 66506, USA,
$^p$University of Manchester, Manchester M13 9PL, United Kingdom,
$^q$Queen Mary, University of London, London, E1 4NS, United Kingdom,
$^r$University of Melbourne, Victoria 3010, Australia,
$^s$Muons, Inc., Batavia, IL 60510, USA,
$^t$Nagasaki Institute of Applied Science, Nagasaki, Japan, 
$^u$National Research Nuclear University, Moscow, Russia,
$^v$University of Notre Dame, Notre Dame, IN 46556, USA,
$^w$Universidad de Oviedo, E-33007 Oviedo, Spain, 
$^x$Texas Tech University, Lubbock, TX  79609, USA,
$^y$Universidad Tecnica Federico Santa Maria, 110v Valparaiso, Chile,
$^z$Yarmouk University, Irbid 211-63, Jordan,
$^{hh}$On leave from J.~Stefan Institute, Ljubljana, Slovenia, 
}}
\noaffiliation

%\noaffiliation
%\collaboration{CDF collaboration}
%%%%%%%%%%%%%%%%%%%%
 \begin{abstract}
     We use a new method to estimate with 5\% accuracy
 the contribution of pion and kaon
 in-flight-decays to the dimuon data set acquired with the CDF detector.
 Based on this improved estimate, we show that
 the total number and the properties of the collected dimuon events are not 
 yet accounted for by ordinary sources of dimuons which also include
 the contributions, as measured in the data, of heavy flavor, $\Upsilon$,
 and Drell-Yan production in addition
 to muons mimicked by hadronic punchthrough.
\end{abstract}
%%%%%%%%%%%%%%%%%%%%
 \preprint{FERMILAB-PUB-11-232-E}
 \maketitle
 %\onecolumn
%%%%%%%%%%%%%%%%%%%%%%%%%%%%%%%%%%%%%%%%%%%%%%%%%%
 \section {Introduction}  \label{sec:ss-intro}
%%%%%%%%%%%%%%%%%%%%%%%%%%%%%%%%%%%%%%%%%%%%%%%%%%

 This article presents an improved determination of the composition
 of a dimuon sample recorded in $p\bar{p}$ collisions at $\sqrt{s}=1.96$ TeV.
  The data sample consists of events
 containing two central ($|\eta|<0.7$) primary (or trigger) muons, each with transverse
 momentum $p_T \geq 3 \; \gevc$, and with invariant mass larger than 
 5 $\gevcc$ and smaller than 80 $\gevcc$.
 The sample may be dominated by real muon pairs due to semileptonic decays of heavy flavor,
 Drell-Yan production and $\Upsilon$ decays, but also contains events in which one or both muons
 are produced by hadrons that decay in flight or otherwise mimic a muon signal.
 Although the dimuon signature can be a powerful tool with which to search for new physics or 
 sources of CP violation,
 the uncertainty of the in-flight-decay contribution makes the precise determination of the fractions of
 known processes a serious experimental challenge. In particular, it remains controversial if
 muons originating from the decay of objects with a lifetime longer than that of heavy-flavored hadrons
 can be completely accounted for with ordinary sources such as in-flight-decays.
 Earlier and recent studies  estimate the fraction of this type of event
 to be negligible~\cite{2mucdf,bmix,williams}.
 Other studies find it significant,  suppress it by selecting muons
 produced close to the beamline~\cite{bbxs}, but have estimated its size with a very large
 uncertainty by using Monte Carlo simulations~\cite{a0disc}. 
 The present work is based on the same Monte Carlo simulated samples, and the
 same analysis methods as Refs.~\cite{bbxs,a0disc},
 but we  improve the method to estimate the number of events due to 
 in-flight-decays achieving a 5\% accuracy.

 Section~\ref{sec:ss-det} describes the CDF~II detector.
 In Sec.~\ref{sec:ss-expsit}, we review the present experimental situation.
 Sections~\ref{sec:ss-anal} to~\ref{sec:ss-fit} describe the procedure used to tune
 the simulation and estimate the contribution of ordinary sources to events
 in which muons are produced by objects with very long lifetimes.
 Based on this results, Section~\ref{sec:ss-multimu} updates the estimate 
 of the rate of multi-muon events reported in Ref.~\cite{a0disc}.
 Our conclusions are presented in Sec~\ref{sec:ss-concl}. 
%%%%%%%%%%%%%%%%%%%%%%%%%%%%%%%%%%%%%%
 \section{CDF II detector and trigger} \label{sec:ss-det}
%%%%%%%%%%%%%%%%%%%%%%%%%%%%%%%
 CDF~II is a multipurpose detector, equipped with a charged particle
 spectrometer and a finely segmented calorimeter. In this section, we 
 describe the detector components that are relevant to this analysis.
 The description of these subsystems can be found in
 Refs.~\cite{det1,det2,det3_0,det3,det4_0,det4,det5,det6,det7,det8}.
 Two devices inside the 1.4 T solenoid are used for measuring the momentum
 of charged particles: the silicon vertex detector (SVXII and ISL) and the
 central tracking chamber (COT). The SVXII detector consists of 
 microstrip sensors arranged in six cylindrical shells with radii between
 1.5 and 10.6 cm, and with a total $z$ coverage~\footnote{
 In the CDF coordinate system, $\theta$ and $\phi$ are the polar and
 azimuthal angles of a track, respectively, defined with respect to the
 proton beam direction, $z$. The pseudorapidity $\eta$ is defined as 
 $-\ln \;\tan (\theta/2)$. The transverse momentum of a particle is 
 $p_T= p \; \sin (\theta)$. The rapidity is defined as 
 $y=1/2 \cdot \ln ( (E+p_z)/(E-p_z) )$, where $E$ and $p_z$ are the 
 energy and longitudinal momentum of the particle associated with the track.} 
 of 90 cm. The first SVXII layer, also referred to as the L00 detector,
 is made of single-sided sensors mounted on the beryllium beam pipe.
 The remaining five SVXII layers are made of double-sided sensors and 
 are divided into three contiguous five-layer sections along the beam 
 direction $z$. The vertex $z$-distribution for $p\bar{p}$ collisions is
 approximately described by a Gaussian function with a rms of 28 cm.
 The transverse profile of the Tevatron beam is circular and has a rms
 spread of $\simeq 25\; \mu$m in the horizontal and vertical directions.
 The SVXII single-hit resolution is approximately $11\; \mu$m and allows
 a track impact parameter
 resolution of approximately $35\; \mu$m, when
 also including the effect of the beam transverse size. The two additional 
 silicon layers of the ISL help to link tracks in the COT to hits in the
 SVXII. The COT is a cylindrical drift chamber containing 96 sense wire
 layers grouped into eight alternating superlayers of axial and stereo
 wires. Its active volume covers $|z| \leq 155$ cm and 40 to 140 cm in
 radius. The transverse momentum resolution of tracks reconstructed using
 COT hits is $\sigma(p_T)/p_T^2 \simeq 0.0017\; [\gevc]^{-1}$. The trajectory
 of COT tracks is extrapolated into the SVXII detector, and tracks are
 refitted with additional silicon hits consistent with the track extrapolation.

 The central muon detector (CMU) is located around the central electromagnetic
 and hadronic calorimeters, which have a thickness of 5.5 interaction lengths
 at normal incidence. The CMU detector covers a nominal pseudorapidity range
 $|\eta| \leq 0.63$ relative to the center of the detector, and  is segmented
 into two barrels of 24 modules, each covering 15$^\circ$ in $\phi$. Every
 module is further segmented  into three  submodules, each covering
 4.2$^\circ$ in $\phi$ and consisting of four layers of drift chambers.
 The smallest drift unit, called a stack, covers a 1.2$^\circ$ angle in
 $\phi$. Adjacent pairs of stacks are combined together into a tower.
 A track segment (hits in two out of four layers of a stack) detected in
 a tower is referred to as a CMU stub. A second set of muon drift chambers
 (CMP) is located behind an additional steel absorber of 3.3 interaction
 lengths. The chambers are 640 cm long and are arranged axially to form a
 box around the central detector. The CMP detector covers a nominal
 pseudorapidity range $|\eta| \leq 0.54$ relative to the center of the
 detector. Muons which produce a stub in both the CMU and CMP systems are
 called CMUP muons. The CMX muon detector consists of eight drift chamber
 layers and scintillation counters positioned behind the hadron calorimeter.
 The CMX detector extends the muon coverage to $|\eta| \leq 1$ relative to
 the center of the detector.

 The luminosity is measured using gaseous Cherenkov counters (CLC) that
 monitor the rate of inelastic $p\bar{p}$ collisions. The inelastic 
 $p\bar{p}$ cross section at $\sqrt{s}=1960$ GeV is scaled from measurements
 at $\sqrt{s}=1800$ GeV using the calculations in Ref.~\cite{sigmatot}.
 The integrated luminosity is determined with a 6\% systematic
 uncertainty~\cite{klimen}.
 
 CDF uses a three-level trigger system. At Level 1 (L1), data from every
 beam crossing are stored in a pipeline capable of buffering data from 42
 beam crossings. The L1 trigger either rejects events or copies them into
 one of the four Level 2 (L2) buffers. Events that pass the L1 and L2 
 selection criteria are sent to the Level 3 (L3) trigger, a cluster of
 computers running  speed-optimized reconstruction code.  

 For this study, we select events with two muon candidates identified by the
 L1 and L2 triggers. The L1 trigger uses tracks with $p_T \geq 1.5 \; \gevc$
 found by a fast track processor (XFT). The XFT examines COT hits from the
 four axial superlayers and provides $r-\phi$ information in azimuthal
 sections of 1.25$^\circ$. The XFT passes the track information to a set of
 extrapolation units that determine the CMU towers in which a CMU stub  
 should be found if the track is a muon. If a stub is found, a L1 CMU
 primitive is generated. The L1 dimuon trigger requires at least two CMU
 primitives, separated by at least two CMU towers. The L2 trigger 
 additionally requires that at least one of the muons also has a CMP stub
 matched to an XFT track with $p_T \geq 3 \;\gevc$. All these trigger
 requirements are emulated by the detector simulation on a run-by-run basis.
 The L3 trigger requires a pair of CMUP muons with invariant mass larger
 than $5 \; \gevcc$, and $|\delta z_0| \leq 5$ cm, where $z_0$ is the $z$
 coordinate of the muon track at its point of closest approach to the beamline
 in the $r-\phi$ plane. These requirements define the dimuon trigger
 used in this analysis.
%%%%%%%%%%%%%%%%%%%%%%%%%%%%%
\section{Present understanding of the dimuon sample composition}
 \label{sec:ss-expsit}
  The value of
 $\sigma_{b\rightarrow\mu,\bar{b}\rightarrow \mu}$
 and  $\sigma_{c\rightarrow\mu,\bar{c}\rightarrow \mu}$, the correlated cross sections
 for producing pairs of central heavy-flavored quarks that decay semileptonically, is
 derived in Ref.~\cite{bbxs} by 
 fitting the
 impact parameter~\cite{d0} distribution of the primary muons with the
 expected shapes from all sources believed to be significant:
 semileptonic heavy flavor decays, prompt
 quarkonia decays, Drell-Yan production, and instrumental backgrounds
 due to punchthrough of prompt or heavy-flavored hadrons which mimic a  muon signal~\cite{strange}.
 In the following, the sum of
 these processes
 will be referred to as the prompt plus heavy flavor ($P+HF$) contribution.
 The notation $K^{puth} \rightarrow \mu$ and
 $\pi^{puth} \rightarrow \mu$ will be used to indicate muon signals mimicked
 by punchthrough of kaons and pions, respectively.
 In order to properly model the data with the templates of the
 various $P+HF$ sources, the study in Ref.~\cite{bbxs} has used strict 
 selection  criteria, referred to as tight SVX selection in the following,
 by requiring muon tracks with hits in the two innermost 
 layers of the SVX
 detector, and in at least two of the next four outer layers.
 
 The tight SVX requirements select events in
 which both muons arise from parent particles that have decayed within
 a distance of $\simeq 1.5$ cm from the  $p\bar{p}$ interaction primary
 vertex in the plane transverse to the beamline. 
 This requirement  suppresses the yield of primary muons due to
 in-flight-decays of pions and kaons, in the following referred to as
  $\pi^{ifd} \rightarrow \mu$ and $K^{ifd} \rightarrow \mu$, respectively.
 This type of contribution to the dimuon dataset prior to any SVX requirement
  was considered negligible
 in previous~\cite{2mucdf,bmix} and  recent~\cite{williams} studies by the CDF and
 D0 collaborations.

\begin{figure}[htp]
 \begin{center}
 \vspace{-0.2in}
 \leavevmode
 \includegraphics*[width=0.5\textwidth]{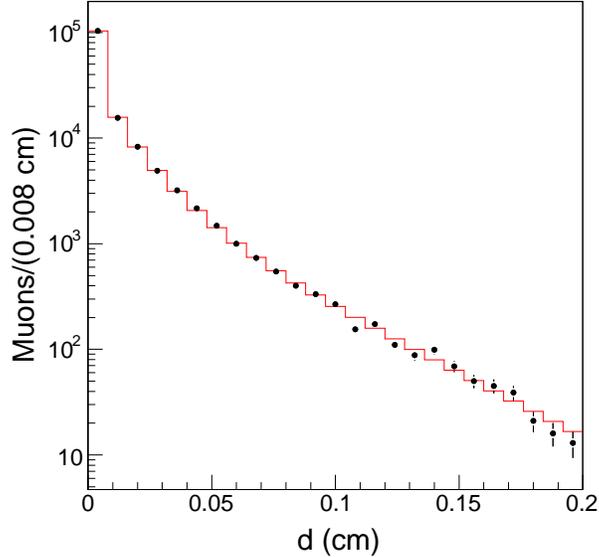}
 \caption[]{The projection of the  two-dimensional impact parameter
            distribution of muon pairs onto one of the two axes is 
            compared to the fit result (histogram).}
 \label{fig:figbb_6}
 \end{center}
 \end{figure}

 As shown by Fig.~\ref{fig:figbb_6}, the tight SVX
 sample is well modeled by fits using the prompt and heavy flavor
 contributions~\cite{bbxs}.
 The sample composition determined by the fit and corrected for the
 appropriate efficiency of the tight SVX requirements~\footnote{
 The efficiency of the tight SVX selection has been measured~\cite{a0disc} to be $0.257 \pm 0.004$ for
 prompt dimuons and $0.237\pm0.001$ for dimuons produced by heavy flavor decays
 by  using control
 samples of data from various sources 
  ($J/\psi\rightarrow \mu^+\mu^-$, $B^\pm \rightarrow \mu^+ \mu^- K^\pm$,
 $B \rightarrow \mu D^0$, and $\Upsilon\rightarrow \mu^+\mu^-$).}
 is listed in
 the first two columns of Table~\ref{tab:tab_1}.
%%%%%%%%%%%%%%%%%%%%%%%%%%%%%%%%%%
 \begin{table}[htp]
 \caption[]{Number of events attributed to the different dimuon sources by
           the fit to the muon impact-parameter distribution.
            The fit parameters $BB$, $CC$, and $PP$ represent 
           the $b\bar{b}$, $c\bar{c}$, and prompt dimuon contributions, 
           respectively. The component $BC$ represents events containing
           $b$ and $c$ quarks. The fit parameter $BP$ ($CP$) estimates
           the number of events in which there is only one $b$ ($c$) quark
           in the detector acceptance and the second muon is produced by  
           prompt hadrons in the recoiling jet that mimic a muon signal.
           Real muons are muons from semileptonic decay of heavy flavors,
           Drell-Yan production or quarkonia decays.
          The data correspond to an
           integrated luminosity of 742 pb$^{-1}$. The dimuon data set consists of
           743006 events.}
 \begin{center}
 \begin{ruledtabular}
 \begin{tabular}{lccc}
%\hline
%\hline
 Component  &  No. of Events & No. of real $\mu-\mu$ & No. and type of misidentified $\mu$     \\
  $BB$      & $230308 \pm 2861$ &  $R_{bb} \times BB $  &
  $ R_{bb} \times [7902 \; K^{puth} \rightarrow \mu $    \\
            &     &              & +8145 $ \pi^{puth} \rightarrow \mu$ ] \\ 
  $CC$      & $103198 \pm 6603$ & $ R_{cc} \times CC$ & 
   $ R_{cc} \times [17546 \; K^{puth} \rightarrow \mu$ \\
            &           &        & + 9535 $\pi^{puth} \rightarrow \mu$] \\
  $PP$      & $161696 \pm 2533$ & $\Upsilon=51680 \pm 649$  & 4400 $K^{puth} \rightarrow \mu\; K^{puth} \rightarrow \mu$  \\
            &                   &   + $DY=54200 \pm 5420$      & +30000 $\pi^{puth} \rightarrow \mu\; \pi^{puth}\rightarrow \mu$  \\
            &                   &                               & + 23000 $K^{puth} \rightarrow \mu\; \pi^{puth} \rightarrow \mu$ \\
 $BP$      & $~43096 \pm  3087$ & &11909  $K^{puth} \rightarrow \mu$ + 29253 $\pi^{puth} \rightarrow \mu $  \\
 $CP$      & $~41582 \pm 5427$ & & 16447  $K ^{puth} \rightarrow \mu$ + 35275 $\pi ^{puth} \rightarrow \mu $     \\
  $BC$      & $~~9135 \pm 2924$  & &   \\
 $P+HF$    & $589015 \pm 5074$ & & \\
%\hline
%\hline
 \end{tabular}
 \end{ruledtabular}
 \end{center}
 \label{tab:tab_1}
 \end{table}
%%%
 %%%%%%%%%%%%%%%%%%%%%%%%%%
%%%%%%%%%%%%%%%%%%%%%%%%%%%%%%%%%%%%%%%%

 The difference between the total number of dimuons and the 
 $P+HF$ component indicates the presence of an important source of dimuons
 produced beyond 1.5 cm  which is
 suppressed by  the tight SVX requirements.
 Because unnoticed by previous experiments, this
 source was whimsically referred to as the ghost contribution.

 The relative size of the ghost and $P+HF$ contributions depends upon the type 
 of SVX requirement applied to the trigger muons.
 Reference~\cite{a0disc} shows that neglecting the presence of ghost events
 affected previous measurements of 
  $\sigma_{b\rightarrow\mu,\bar{b}\rightarrow \mu}$~\cite{2mucdf,d0b2}
 and of  $\bar{\chi}$~\cite {bmix} at the Tevatron.
 Finally, the ghost sample is shown 
 to be the source of the dimuon invariant mass discrepancy 
 observed in Ref.~\cite{dilb}. 
 
 Reference~\cite{a0disc} has studied a number of potential sources  
 of muons  originating beyond the beam pipe.
 Contrary to what assumed by previous experiments, the one source 
 found to contribute significantly arises
 from in-flight-decays of pions and kaons.
 Based upon a generic QCD simulation,
 that study estimates a contribution of 57000 events.
 A smaller 
 contribution ($ 12052\pm 466$ events) from $K^0_S$ and hyperon decays in
 which the punchthrough of a hadronic prong mimics a muon signal was estimated
 using the data. Secondary inelastic interactions in the tracking volume were
 found to be a negligible source of ghost events.
 The final estimate  of the size of possible sources of ghost events
 underpredicts
 the observed number by approximately a factor of two (154000
 observed and 69000 accounted for), but the
 difference was not considered significant because 
 of the simulation uncertainty. 
 
 The present study uses events selected with the tight SVX requirements to tune the
 QCD simulation. Since these data are well modeled by the impact parameter templates of the
 $P+HF$ components, misidentified muons can only arise from the punchthrough of prompt hadrons or
 hadrons produced by heavy-flavor decays. The numbers of misidentified muons
 in the data are derived
 by subtracting the expected number of real muons, listed in the third
 column of Table~\ref{tab:tab_1},  from the corresponding
 components in the second column. We then compare
 these differences to the rate of  $K^{puth} \rightarrow \mu$ and
 $\pi^{puth} \rightarrow \mu$ misidentifications predicted by the simulation, and
 listed on the fourth column of the same table.
 The simulation is tuned by adjusting the predicted rate of pions and kaons
 to reproduce the observed number of
  muon misidentifications. Then, the tuned simulation
 is used to predict the number of muons due to in-flight-decays with 5\% accuracy.
%%%%%%%%%%%%%%%%%%%%%%%%%%%%%%%
 \section{Rates of misidentified muons in the data and simulation} \label{sec:ss-anal}
%%%%%%%%%%%%%%%%%%%%%%%%%%%%%%%
 
 We make use of three different samples of simulated events
 generated with the {\sc herwig} parton-shower Monte-Carlo program~\cite{herwig},
 the settings of which are described in Appendix~A
 of Ref.~\cite{bbxs}.
  We use option 1500 of the {\sc herwig} program to generate final states
 produced by hard scattering of partons with transverse momentum larger
 than 3 $\gevc$ (sample A=generic QCD).
 Hadrons with heavy flavors  are subsequently 
 decayed using the {\sc evtgen} Monte Carlo program~\cite{evtgen}. 
 The detector response to particles produced by the above generators
 is modeled with the CDF~II detector simulation that in turn is based on 
 the {\sc geant} Monte Carlo program~\cite{geant}.
 The values of the heavy flavor cross sections
 predicted by the generator are scaled to the measured values
  $\sigma_{b\rightarrow\mu,\bar{b}\rightarrow \mu}= 1549 \pm 133$ pb
 and  $\sigma_{c\rightarrow\mu,\bar{c}\rightarrow \mu}= 624 \pm 104$ pb~\cite{bbxs}. 
 The  next simulated sample (sample B=single $b$ + single $c$) is extracted from A by 
requiring the presence of at least a
 trigger muon generated from heavy-flavor semileptonic decays.
  The simulated sample C=$b\bar{b}$ + $c\bar{c}$ is extracted from B by requiring the presence of at least two
 trigger muons generated from heavy flavor decays. This sample has been used
 to construct  impact parameter templates and estimate 
 kinematic acceptances in Ref.~\cite{bbxs}.

 In the various simulations, we evaluate the number of dimuons from heavy flavor decays
 and the number of pairs of tracks of different type that pass the same kinematic selection.
 The ratios of these numbers are listed in Tables~\ref{tab:tab_2} to~\ref{tab:tab_4}.
 The rate of pairs of tracks of different type predicted by the simulation are normalized to the data by
 multiplying these ratios by  the number of dimuons from $b\bar{b}$ or $c\bar{c}$ production
 observed in the data. 
 %%%%%%%%%%%%%%%%%%%%%%%%%%%%%%%%%%
 \begin{table}[htp]
 \caption[]{Ratio of the numbers of $\pi\pi$, $KK$, and $K\pi$ pairs to that of
            primary dimuons from $b\bar{b}$ decays (221096 pairs) in the generic QCD
        simulation (sample A).}
 \begin{center}
 \begin{ruledtabular}
 \begin{tabular}{lccc}
% \hline
%\hline 
  Process& $R_{KK}$ &  $R_{K\pi}$ &  $R_{\pi\pi}$  \\
  generic QCD &  867 &  8935   &   22913  \\ 
 %\hline
%\hline 
 \end{tabular}
 \end{ruledtabular}
 \end{center}
 \label{tab:tab_2}
 \end{table}

%%%%%%%%%%%%%%%%%%%%%%%%%%%%%%%%%%%
 %%%%%%%%%%%%%%%%%%%%%%%%%%%%%%%%%%
 \begin{table}[htp]
 \caption[]{Ratio of the numbers of $\mu-K(\pi)$ combinations to that of
          primary dimuons from $b\bar{b}$ and $c\bar{c}$ production in the single-$b$ 
          and single-$c$ simulated samples (221096 and 83590 dimuons, respectively).}
 \begin{center}
 \begin{ruledtabular}
 \begin{tabular}{lcc}
% \hline
%\hline 
  Process  &  $R_K$ &  $R_\pi$     \\
  single $b$  & 11.1  &  ~54.4   \\ 
  single $c$  & 40.7  & 173.7      \\
% \hline
%\hline 
 \end{tabular}
 \end{ruledtabular}
 \end{center}
 \label{tab:tab_3}
 \end{table}

%%%%%%%%%%%%%%%%%%%%%%%%%%%%%%%%%%%%%%%%%%%
 %%%%%%%%%%%%%%%%%%%%%%%%%%%%%%%%%%
 \begin{table}[htp]
 \caption[]{Ratio of the numbers of $\mu-K(\pi)$ combinations to that of
          primary dimuons from heavy flavor production in the $b\bar{b}$ 
          and $c\bar{c}$ simulated samples (221096 and 83590 dimuons, respectively).}
 
 \begin{center}
 \begin{ruledtabular}
 \begin{tabular}{lcc}
 %\hline
%\hline 
  Process   &  $R_K$ &  $R_\pi$     \\
  $b\bar{b}$   & ~7.4  &  15.2   \\ 
  $c\bar{c}$   & 43.5 & 46.9       \\
% \hline
%\hline 
 \end{tabular}
 \end{ruledtabular}
 \end{center}
 \label{tab:tab_4}
 \end{table}
%%%%%%%%%%%%%%%%%%%%%%%%%%%%%%%%%%%%%%%%%%%

 The probability  $P^{puth}_{K(\pi)}$ that a kaon (pion) is not contained by the calorimeter
 and mimics a muon signal 
 has been measured in Ref.~\cite{bbxs} by using
 kaons and pions from $D^{*\pm} \rightarrow \pi^\pm  D^0$ with $D^0 \rightarrow K^+ \pi^-$ decays.
 The probability that  kaon (pion) in-flight-decays  mimic a trigger
 muon,  $P^{ifd}_{K(\pi)}$, has been derived in Ref.~\cite{a0disc} by using 
 the simulated sample C.  These probabilities depend on the particle transverse momentum.
 Table~\ref{tab:tab_5} lists the average probabilities that kaons (pions) mimic
 a primary muon when applying the $P^{puth}_{K(\pi)}$ and  $P^{ifd}_{K(\pi)}$
 probabilities to
 simulated  kaon (pion) tracks
 with $p_T \geq 3 \; \gevc$ and $|\eta|<0.7$.
%%%%%%%%%%%%%%%%%%%%%%%%%%%%%%%%
 \begin{table}[htp]
 \caption[]{ Average probabilities (\%) that  punchthroughs or in-flight decays result into
 a primary muon. The $p_T$ distribution of kaons and pions in the different simulations are
 almost indistinguishable.}
 \begin{center}
 \begin{ruledtabular}
 \begin{tabular}{cccc}
 %\hline
%\hline 
  $<P^{puth}_K> $ & $<P^{puth}_{\pi}>$&  $<P^{ifd}_K>$ & $<P^{ifd}_\pi>$ \\
    $0.483\pm 0.003$ &   $0.243 \pm.0.004$ & $0.345\pm 0.005$  & $0.0727\pm 0.0016$ \\
% \hline
%\hline  
\end{tabular}
 \end{ruledtabular}
 \end{center}
 \label{tab:tab_5}
 \end{table}
%%%%%%%%%%%%%%%%%%%%%%%%

 By weighting  simulated pion (kaon) tracks that
 pass the muon kinematic selection with the corresponding
   $P^{puth}_{K(\pi)}$ probability, we obtain the prediction
 of misidentified primary muons for the various $P+HF$ components that is
 listed in the fourth column of Table~\ref{tab:tab_1}.
 The third column of the same table lists the number of real muons for the various
  $P+HF$ contributions. 
 The sum of real plus misidentified muon pairs is in general agreement 
 with the data listed in the second column of the table.
 Therefore, it is reasonable to use the observed rate of dimuons, the
 knowledge of the fraction of real dimuons  due to semileptonic decay of heavy flavors,
 Drell-Yan or $\Upsilon$ mesons, and the knowledge of the $P^{puth}_{K(\pi)}$ probabilities
 to normalize the absolute yields of  pions and kaons 
 predicted by the simulation. The simulation fitted to the data is then  used to predict
 the rate of events due to in-flight-decay misidentifications
  by weighting simulated tracks with the $P^{ifd}_{K(\pi)}$ probabilities,
 the average of which is listed in Table~\ref{tab:tab_5}.
  In addition, the total rate of $K \rightarrow \mu= K^{puth} \rightarrow \mu + K^{ifd} \rightarrow \mu$
 misidentifications predicted by the simulation
 can be further constrained with data. 
 This is done in the next section
 by using the number of primary muons
 due to misidentification of $K^{*0}$, $K^{*\pm}$, and
 $K^0_S$ decays. 

 We first describe the evaluation of the content of real muons in the various $P+HF$ components
 and the  function used to fit the simulation to the data. 
 Reference~\cite{bbxs}
 estimates that the fraction $R_{bb}=0.96 \pm 0.04$ of the $BB$
 component is due to real muons from $b$-quark semileptonic decays whereas
 the remaining 4\% is due to muons mimicked by the punchthrough of hadrons produced by
 heavy flavor decays.
 Similarly, the fraction $R_{cc}=0.81 \pm 0.09$ of the $CC$
 component is due to real muons from $c$-quark semileptonic decays whereas
 the remaining 19\% is due to muons mimicked by
 the punchthrough of hadrons produced by heavy flavor decays.
 The uncertainty of the fraction of real muons due to $b\bar{b}$ ($c\bar{c}$)
 production is accounted for by multiplying  $R_{bb}$ ($R_{cc}$) by
 the fit parameter $f_{bb}$ ($f_{cc}$) constrained to 1 with a 4\% (11\%) Gaussian error.

 The number of $\Upsilon$ mesons contributing to the $PP$ component ($\Upsilon=51680 \pm 649$ candidates)
 has been determined
 in Ref.~\cite{bbxs} by fitting the dimuon invariant mass spectrum with
 three Gaussian functions to model the signal
 and a straight line to model the combinatorial background. The Drell-Yan contribution is evaluated as
 $DY=\Upsilon \times \sigma_{DY}/\sigma_{\Upsilon}$. The cross section $\sigma_{DY}$ 
 in the $5-80\, \gevcc$ mass range is evaluated with a NLO calculation~\cite{drelly}, and we use
 the measured value of  $\sigma_{\Upsilon}$~\cite{yxs}.
 The ratio $\sigma_{DY}/\sigma_{\Upsilon}$ is 1.05 with a 10\% error mostly due to the
 measurement in Ref.~\cite{yxs}. 
 To account for the uncertainty, we weight the $DY$ contribution with the fit parameter
 $f_{dy}$ constrained to 1 with a 10\% Gaussian error.

  The magnitude of the $BP$  ($CP$) component, predicted with the single-$b$ (single-$c$)
 simulation, with respect to that of the $BB$ ($CC$) contribution
  depends on the ratio of NLO to LO 
 terms evaluated by the {\sc herwig} generator. Because of the dependence on the renormalization
 and factorization scales, the uncertainty of the single-$b\;(c)$ cross section to that of the
 $b\bar{b}\; (c\bar{c})$ cross section is estimated~\cite{mnr} to be $\simeq$ 20 (30)\%~\footnote{
  However, the study in Ref.~\cite{ajets} shows that  the {\sc herwig} generator
 predicts the observed single and correlated
 heavy-flavor cross sections to better than 10\%.}.   
 We account for this uncertainty by
 weighting  the rate of pion and kaon tracks predicted by the single-$b$ (single-$c$)
 simulation  with the additional fit parameter $f_{sb}$ ($f_{sc}$)
 constrained to 1 with a 20\% (30\%) Gaussian error.

 The simulation prediction of the number of  muons mimicked by the punchthrough of pions (kaons)
 is weighted with the fit free parameter $f_{\pi}$  ($f_{K}$). 
 These fit parameters provide the absolute normalization of
  the  pion (kaon) rate predicted by the simulation
 including the uncertainties of the punchthrough probabilities.
\section{Measurement of the $K \rightarrow \mu$ contribution} 
\label{sec:ss-tkaon}
 The small   rate of $K \rightarrow \mu = K^{puth} \rightarrow \mu + K^{ifd} \rightarrow \mu$
 misidentifications is measured using a higher
statistics sample of dimuon events corresponding to an  integrated luminosity of 3.9 fb$^{-1}$.
 The number of $K \rightarrow \mu$ misidentification, $N_K$,
  is derived from $N_{K^{*0}}$,  the number of identified  $K^{*0} \rightarrow K^+\pi^-$ decays
  with $K^+\rightarrow \mu^+$ (and  charge-conjugate states).
  The number  $N_{K^{*0}}$ is related to $N_K$ by 
 $$N_{K^{*0}}=N_K \cdot \epsilon_0 \cdot R(K^{*0}),$$
 where $R(K^{*0})$ is the fraction of kaons that result from $K^{*0} \rightarrow K^+\pi^-$ decays
 and $\epsilon_0$ is the efficiency to reconstruct the pion.  

 We also select $K^0_S \rightarrow \pi^+ \pi^-$ with $\pi \rightarrow \mu$
 candidates and reconstruct  $K^{*\pm} \rightarrow K^0_S \pi^\pm$ decays.
 The number of $K^{*\pm}$ is related to that of $K^0_S$ by
 $$N_{K^{*\pm}}=N_{K^0_S} \cdot \epsilon_1 \cdot R(K^{*\pm}),$$
 where  $R(K^{*\pm})$ is the fraction of $K^0_S$ resulting from $K^{*\pm} \rightarrow K^0_S \pi^\pm$
 decays and $\epsilon_1$ is the efficiency to reconstruct the additional pion. 
 We use isospin invariance to set $R(K^{*\pm})= R(K^{*0})$. Since the additional pion 
 used to search for the $K^{*\pm}$ and  $K^{*0}$ candidates is selected with
  the same kinematic requirements, we set
 $\epsilon_0=\epsilon_1$. It follows that
  $$N_K = N_{K^0_S}/N_{K^{*\pm}} \times N_{K^{*0}}.$$

\begin{figure}[htp]
 \begin{center}
 \vspace{-0.2in}
 \leavevmode
 %\epsfxsize \textwidth
 % plot is S2 in plot_note.root
 \includegraphics*[width=0.5\textwidth]{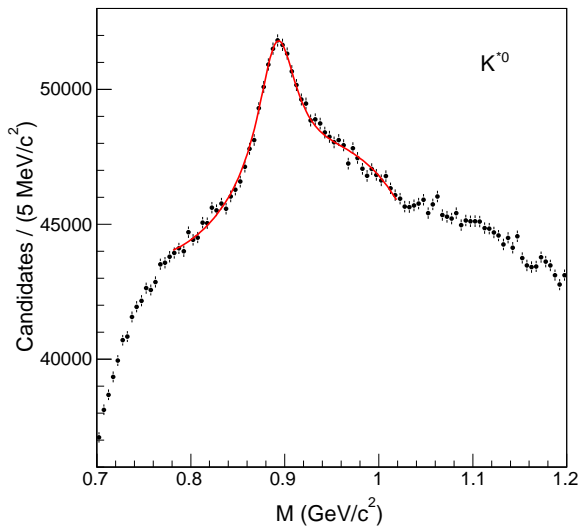}
 \caption[]{Invariant mass distribution of $K^{*0} \rightarrow K^+\pi^-$ candidates
           passing our selection criteria. The line represents the fit described in the text.}
 \label{fig:fig_2}
 \end{center}
 \end{figure}

 We search for $K^{*0}$ decays
 by combining primary muons, assumed to be kaons, with all opposite charge tracks,
 assumed to be pions, with $p_T \geq 0.5 \; \gevc$
 and in an angular cone with $\cos \theta \geq 0.6$ around
 the direction of the primary muon.
 We require tracks with at least 10 axial and 10 stereo COT hits.
 We constrain the pair to arise from a common three-dimensional point, and
 reject combinations if the probability of the vertex-constrained fit is smaller than 0.001.
 The invariant mass spectrum of the selected $K^{*0} \rightarrow K^+\pi^-$ 
 candidates is shown in Fig.~\ref{fig:fig_2}.
 %%%%%%%%%%%%%%%%%%%%%%%%%%
%%%%%%%%%%%%%%%%%%%%%%%%%%%%%%%%%%%%%%%%
 We fit the invariant mass distribution with a Breit-Wigner function smeared by the
 detector mass resolution to model the signal.
  We fix the mass and width
 of the Breit-Wigner function to 896 and 51 MeV/$c^2$~\cite{pdg}, respectively~\footnote{
 The mass resolution due to the track reconstruction in
 simulated events which include kaon in-flight-decays
 is 4.9 MeV/$c^2$, and is negligible compared to the resonance width.}.
 We use a fourth order polynomial to model the combinatorial background under
 the signal, and the fitted range of invariant mass is
 conveniently chosen to yield a fit with 50\% probability.
 The size of the signal is not affected by the  arbitrary 
 choice of the function used to model the combinatorial
 background or of the fitted mass range, and is solely determined by the accurate knowledge of the
 signal shape.
 The fit yields  $N_{K^{*0}}=87471 \pm 2217$ $K^{*0}$ mesons.

\begin{figure}[htp]
 \begin{center}
 \vspace{-0.2in}
 \leavevmode
 %\epsfxsize \textwidth
 % S1 in plot_note.root 
\includegraphics*[width=0.5\textwidth]{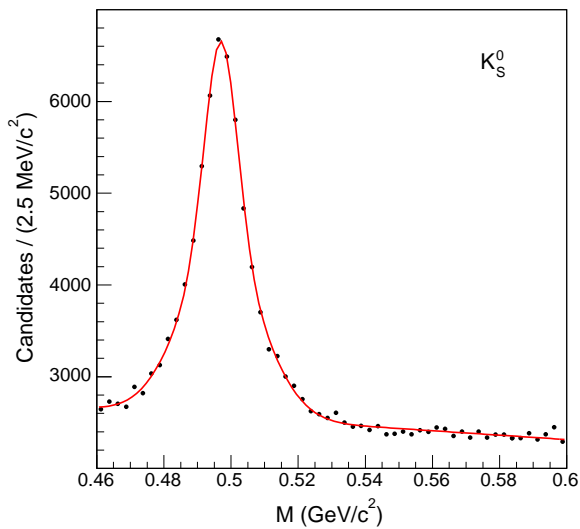}
 \caption[]{Invariant mass distribution of $K^0_S$ candidates
           passing our selection criteria. The line represents the fit described in the text.}
 \label{fig:fig_3}
 \end{center}
 \end{figure}

 We search for
 $K^0_S \rightarrow \pi^+ \pi^-$ with a $\pi \rightarrow \mu$ misidentification
  by combining primary muons
 with tracks passing the same requirements as those used in the $K^{*0}$ search.
 In this case, both tracks are assumed to be pions.
 As in the previous case, we select pairs consistent with arising from a common three-dimensional vertex.
 We take advantage of the $K^0_S$ long lifetime to suppress the combinatorial background. 
 We further require that the distance between the $K^0_S$ vertex and the event primary vertex,
 corrected by the $K^0_S$ Lorentz boost, corresponds to  $ct>0.1$ cm.
 The invariant mass of the $K^0_S$ candidates is shown in  Fig.~\ref{fig:fig_3}.
 %%%%%%%%%%%%%%%%%%%%%%%%%%
%%%%%%%%%%%%%%%%%%%%%%%%%%%%%%%%%%%%%%%%

 We fit the signal with two Gaussian functions and the combinatorial background with a straight line
 in the mass range $0.4-0.6\; \gevcc$.
 Having fixed the peak of the Gaussian functions at 0.497 $\gevcc$~\cite{pdg}, the fit returns an averaged
 $\sigma$ of 8.4 MeV/$c^2$, consistent with what is expected from simulated events~\footnote{
 Because of the $K^0_S$ long decay path, reconstructed track segments may be shorter than the
 available tracking  detector length. When  $K^0_S$ mesons  decay before entering the
 COT volume, the mass resolution is 4  MeV/$c^2$.},
  and a signal of $32445 \pm 421$  $K^0_S$ mesons
 in the mass range $0.474-0.522 \; \gevcc$.

\begin{figure}[htp]
 \begin{center}
 \vspace{-0.2in}
 \leavevmode
 \includegraphics*[width=0.5\textwidth]{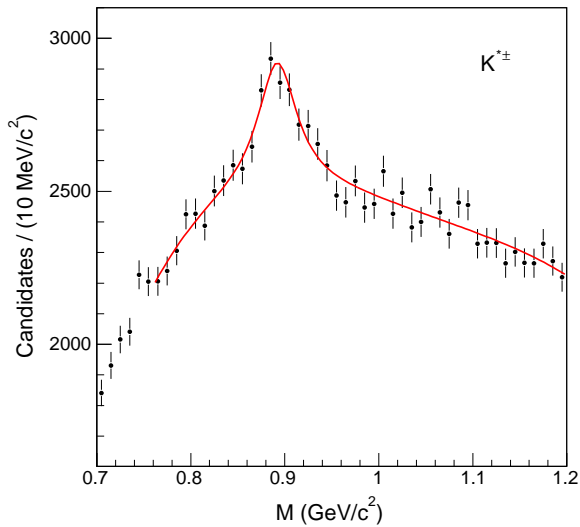}
 \caption[]{Invariant mass distribution of $K^{*\pm} \rightarrow  K^0_S \pi^\pm$ candidates
           passing our selection criteria. The line represents the fit described in the text.}
 \label{fig:fig_4}
 \end{center}
 \end{figure}

 We search for $K^{*\pm}$ by combining $K^0_S$ candidates with mass between 0.474 and 0.522 $\gevcc$
 and   $ct>0.1$ cm with any additional track, assumed to be a pion, that pass the same selection as
 pion tracks used to find $K^{*0}$ candidates. We constrain the $K^0_S$ mass to 0.497 $\gevcc$ and
 require that the   $K^0_S$ candidate and the pion track are consistent with
 arising from a common three-dimensional vertex. 
 The invariant mass distribution of   $K^{*\pm}$ candidates is shown in  Fig.~\ref{fig:fig_4}.
 %%%%%%%%%%%%%%%%%%%%%%%%%%
%%%%%%%%%%%%%%%%%%%%%%%%%%%%%%%%%%%%%%%%
  
  We fit the invariant mass distribution with a Breit-Wigner function to model the signal
 and a fourth order polynomial to model the combinatorial background.  We fix the mass and width
 of the Breit-Wigner function to 892 and 51 MeV/$c^2$~\cite{pdg}, respectively~\footnote{
 In simulated events, when constraining the $K^0_S$ mass to the PDG value, the
 mass-constrained $K^0_S$ momentum is measured as accurately as that of
 a track corresponding to a $K \rightarrow \mu$ decay.
 The  resulting $K^{*\pm}$  mass resolution is approximately 5 MeV/$c^2$.}.  
 The fit returns a signal of $3326 \pm 246$ $K^{*\pm}$ mesons.

 The signals obtained by analyzing the 3.9 fb$^{-1}$ sample are rescaled
to estimate the number of $K \rightarrow \mu$ misidentifications present in the 742 pb$^{-1}$
dataset. After rescaling, 
 we obtain $N_K = N_{K^0_S}/N_{K^{*\pm}} \times N_{K^{*0}}=164769\pm 13067$.
 This number is used to constrain the total
  number $N_K^{\rm sim}$ of $ K \rightarrow \mu$ misidentifications 
 predicted by the simulation.
\section{Fit of the simulation prediction to the data}
\label{sec:ss-fit}
 We fit the simulation prediction with a $\chi^2$-minimization method~\cite{minuit}.
 The   $\chi^2$ function is defined as
 $$\chi^2= \sum_{i=1}^{5} (D[i] -P[i])^2/ED[i]^2 + (N_K^{\rm sim} -164769)^2/13067^2\; ,$$
 where $D[i]$ and $ED[i]$ are the size and error of the component listed in the second column
 and i-th row of Table~\ref{tab:tab_1}. 
  The term $P[i]$ is the sum of the real muon contribution  and the
 punchthrough  contribution predicted by the simulation in the third and forth columns
 of the same table, respectively.
 These contributions are weighted with the fit parameters described in the previous section,
 and the terms read
 
 $$ P[1]= f_{bb}\; (0.96 \cdot 230308 + f_k \; 7902 + f_\pi \; 8145),$$
 $$ P[2]=  f_{cc}\; (0.81 \cdot 103198 + f_k \; 17546 + f_\pi \; 9535),$$
 $$ P[3]= \Upsilon + f_{dy}  \; DY +  f_K^2 \; 4400 + f_\pi^2 \; 30000 +f_K \; f_\pi \; 23000, $$
 $$P[4]=f_{sb}\; (f_K \; 11909 +f_\pi \; 29253),\; {\rm and}$$
  $$P[5]=f_{sc}\; (f_K \; 16447 +f_\pi \; 35275).$$ 
 In addition, the sum  $\sum_{i=1}^{5} D[i]$
 is constrained to the observed number of $P+HF-BC$
 events within its error.
 The fit results are shown in Tables~\ref{tab:tab_6} to~\ref{tab:tab_8}.
 In Table~\ref{tab:tab_6}, the fit parameters that tune the 
 various cross sections predicted by the {\sc herwig} generator
 are very close to their nominal values indicating that the 
 default simulation provides a quite accurate modeling of the data.
%%%%%%%%%%%%%%%%%%%%%%%%%%%%%%%%%%
 \begin{table}[htp]
 \caption[]{Parameter values returned by the fit described in the text.
 The fit yields $\chi^2=2.5$ for 5 DOF.}

\begin{center}
\begin{ruledtabular}
 \begin{tabular}{lc}
% \hline
%\hline 
  $f_{bb}$      &  $0.97 \pm 0.01$     \\ 
  $f_{cc}$      & $ 0.95 \pm 0.04$    \\
  $f_{dy}$      & $  1.01 \pm 0.09$    \\
  $f_\pi$      & $  0.97 \pm 0.08$    \\
  $f_K$      & $  1.01 \pm 0.08$    \\
  $f_{sb}$      & $1.05 \pm 0.08$    \\
  $f_{sc}$      & $0.82 \pm 0.10$    \\
% \hline
%\hline  
\end{tabular}
 \end{ruledtabular}
\end{center}
 \label{tab:tab_6}
 \end{table}
%%%%%%%%%%%%%%%%%%%%%%%%%%%%%%%%%%%%%%%%%%%

\begin{figure}[htp]
 \begin{center}
 \vspace{-0.2in}
 \leavevmode
 %\epsfxsize \textwidth
 % SVX in plot_svx.root
\includegraphics*[width=0.5\textwidth]{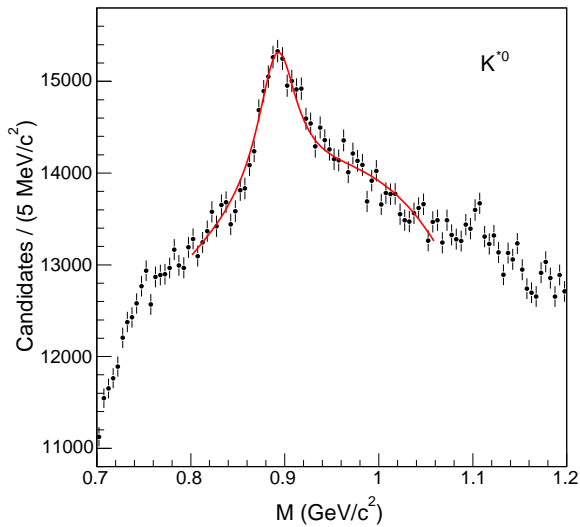}
  \caption[]{Invariant mass distribution of $K^{*0}$ candidates
           in which the $K\rightarrow \mu$ misidentification
 passes the tight SVX requirement. The line represents the fit described in the text.}
 \label{fig:fig_5}

 \end{center}
 \end{figure}

 %%%%%%%%%%%%%%%%%%%%%%%%%%%%%%%%%%%%%%%%%%%
 \begin{table}[htp]
 \caption[]{Parameter correlation coefficients returned by the fit.}
\begin{center}
\begin{ruledtabular}
 \begin{tabular}{lcccccc}
% \hline
%\hline 
  Fit parameter & $f_{bb}$ & $f_{cc}$ & $f_{dy}$ & $f_{\pi}$ & $f_{K}$ &  $f_{sb}$      \\
  $f_{cc}$      & $-0.11$  &          &         &          &         &                  \\
  $f_{dy}$      & $~0.17$  & $~0.10$  &         &          &         &                   \\
  $f_{\pi} $    & $-0.16$  & $-0.10$  & $-0.84$ &          &         &                    \\
  $ f_{K}$      & $-0.08$  & $-0.17$  & $~0.71$ & $-0.48$  &         &                      \\
  $f_{sb} $     & $-0.07$  & $-0.03$  & $~0.50$ & $-0.51$  & $-0.01$ &                    \\
  $f_{sc} $     & $-0.02$  & $-0.21$  & $~0.36$ & $-0.35$  & $-0.12$ & $0.15$  \\
% \hline
%\hline 
 \end{tabular}
 \end{ruledtabular}
 \end{center}
 \label{tab:tab_7}
 \end{table}
%%%%%%%%%%%%%%%%%%%%%%%%%%%%%%%%%%%%%%%%%%%
%%%%%%%%%%%%%%%%%%%%%%%%%%%%%%%%%%
 \begin{table}[htp]
 \caption[]{Number of events due to different production mechanisms 
            are compared to the result of the present fit.}
 \begin{center}
 \begin{ruledtabular}
 \begin{tabular}{lcc}
 %\hline
%\hline 
 Component  &  No. of Events & Fit result    \\
  $BB$      & $230308 \pm 2861$  & 230607  \\
  $CC$      & $103198 \pm 6603$  & 104463 \\ 
  $PP$      & $161696 \pm 2533$  & 161387\\
  $BP$      & $~43096 \pm  3087$ &  42490  \\
  $CP$      & $~41582 \pm 5427$  &  41822   \\
  $P+HF$    & $589015 \pm 5074$ &  589905 \\
 %\hline
%\hline  
\end{tabular}
 \end{ruledtabular}
 \end{center}
 \label{tab:tab_8}
 \end{table}
%%%
 %%%%%%%%%%%%%%%%%%%%%%%%%%
 The fit returns 163501 $K \rightarrow \mu$  candidates ($164769 \pm 13067$ are measured in the data),
  51\% of which are due to punchthrough and 49\%
 to in-flight-decays. We verify this result by measuring the fraction of
 $K\rightarrow \mu$ decays in identified $K^{*0} \rightarrow K^+\pi^-$ decays
 that pass the tight SVX requirements.
 The efficiencies of the tight SVX requirement applied to primary muons
 are $0.356 \pm 0.002$ for muons due to punchthrough of
  prompt and heavy-flavored hadrons and $0.166 \pm 0.005$ for
 muons arising from in-flight decays~\footnote{
 In Ref.~\cite{a0disc}, which uses 0.74 fb$^{-1}$ of data,
 these efficiencies have been measured to be 0.45 and 0.21,
 respectively. In 3.9 fb$^{-1}$ of data, by using $\Upsilon$ candidates, we
 measure a smaller efficiency of the tight SVX selection.
 The efficiency loss comes from periods of data taking in which the pedestals
 of the L00 channels were miscalibrated.
}.
  Based on the kaon composition
 returned by the fit, we estimate the efficiency of the tight SVX requirement
 applied to $K\rightarrow \mu$ misidentifications to be $0.263 \pm 0.008$,
 where the error includes the uncertainty of the efficiencies and that of the
 kaon composition returned by the fit.
 Figure~\ref{fig:fig_5} shows the invariant mass distribution of  $K^{*0}$ candidates
 after applying the tight SVX requirement. 
 %%%%%%%%%%%%%%%%%%%%%%%%%%
%%%%%%%%%%%%%%%%%%%%%%%%%%%%%%%%%%%%%%%% 
 We fit the invariant mass distribution with
 the same function used to fit the $K^{*0}$ mass distribution in Fig.~\ref{fig:fig_2}.
 The fit returns  $22689 \pm 985$   $K^{*0}$ candidates to be
 compared with  $87741 \pm 2219$ $K^{*0}$ candidates
 before applying the tight SVX requirement. The resulting
 efficiency of the tight SVX requirement is $0.253 \pm 0.013$, in agreement
 with what expected ($0.263 \pm 0.008$) using the
 composition of the kaon sample returned by the fit. 
 
 Continuing with the analysis of the results returned by the fit,
 the total number of   $\pi \rightarrow \mu$ misidentifications  is 240915, 64\% of which
 are due to punchthrough and 36\% to in-flight-decays.
 The fractional composition of the  $K \rightarrow \mu$ and  $\pi \rightarrow \mu$
 misidentifications is summarized in Table~\ref{tab:tab_9}.
%%%%%%%%%%%%%%%%%%%%%%%%%%%%%%%%%%
 \begin{table}[htp]
 \caption[]{Contributions (\%) of various processes to pion or kaon
 misidentifications.}
 \begin{center}
 \begin{ruledtabular}
 \begin{tabular}{lcc}
 %\hline
%\hline  
 Type  &  $K^{puth} \rightarrow \mu$ &  $\pi^{puth} \rightarrow \mu$  \\
  All sample                  & 51  & 64  \\
  generic QCD                  & 20  & 33  \\
  single $b$ + $b\bar{b}$ +      &    &  \\ 
  single $c$ + $c\bar{c}$      & 31   & 31\\
 Type  &  $K^{ifd} \rightarrow \mu$ &  $\pi^{ifd} \rightarrow \mu$  \\
 All sample                  & 49  & 36 \\
  generic QCD                  & 27  & 27  \\
  single $b$ + $b\bar{b}$ +      &    &  \\ 
  single $c$ + $c\bar{c}$      & 22   & 9\\
% \hline
%\hline  
\end{tabular}
 \end{ruledtabular}
 \end{center}
 \label{tab:tab_9}
 \end{table}
%%%
 %%%%%%%%%%%%%%%%%%%%%%%%%%
 
 The total fraction of misidentified muons in the dataset is 27\%.
 The number of misidentified muons due in-flight-decays of pions and kaons (ghost events)
 is  $113613 \pm 5332$.
 Since the number of muons from in-flight-decays is derived from that of muons mimicked by
 hadron punchthrough using the fake probabilities listed in Table~\ref{tab:tab_5}, the
 uncertainty of these probabilities yields an additional error of 3845 events.

% It is interesting to note that the fit result is quite insensitive
% to the $N_{K^{*0}}$ error. If we artificially double it, the fit decreases by 2\%
%  the number of $K \rightarrow \mu$ misidentifications and yields  $112680 \pm 5329$
% ghost events. If we triple the error, the fit decreases by 5\%
%  the number of $K \rightarrow \mu$ misidentifications and yields  $111209 \pm 5326$
% ghost events. 
 After adding the $ 12052\pm 466$ events from $K^0_S$ and hyperon decays, we predict
 $125665 \pm 5351$  ghost events, whereas the dimuon dataset
 contains $153991 \pm 5074$  events of this type. The number of
 unaccounted events ($28326 \pm 7374$) is
 ($12.8 \pm 3.2$)\% of the $b\bar{b}$ production and $(18.3 \pm 4.7)$\% of the ghost
 sample.
 
\section{Revised estimate of the rate of additional real muons in ghost events}
\label{sec:ss-multimu}
 As a cross-check of its $b\bar{b}$ content,
 Reference~\cite{a0disc} has investigated the rate of sequential semileptonic
 decays of single $b$ quarks in the dimuon sample.
  We provide here a summary of that study and its conclusions.
 That study searches for  additional muons with $p_T \geq 2 \; \gevc$ and $|\eta| \leq 1.1$
  in a dimuon sample corresponding to an integrated
 luminosity of 1426 pb$^{-1}$. The sample of 1426571 events
 contains $1131090 \pm 9271$  $P+HF$ events  in which both muons originate inside the beam pipe
 and $295481 \pm 9271$ ghost events in which at least one muon is produced outside.
 The study selects  pairs of primary and additional muons with opposite charge ($OS$) and invariant mass
 smaller than 5 $\gevcc$.

 In the case of Drell-Yan or quarkonia production, which was not simulated, the rate of same-charge pairs ($SS$)
 is a measure of the fake muon contribution since misidentified muons arise from the underlying event
 which has no charge correlation with primary muons.
 The rate of additional muons mimicked by hadronic punchthrough is also estimated with a 
 probability per track derived by using
 kaons and pions from $D^{*\pm} \rightarrow \pi^\pm  D^0$ with $D^0 \rightarrow K^+ \pi^-$ decays.
 This misidentification probability  is approximately
 ten times larger than that for primary muons that have to penetrate twice as many
 interaction lengths~\footnote{ Therefore, the contribution of in-flight-decays
 to additional muons is negligible in comparison with the punchthrough contribution.}.
 The punchthrough probabilities for pions and kaons differ by a factor of two.
 In addition, in simulated  events due to heavy flavor production,
 the pion to kaon ratio depends on the invariant mass and the
 charge of the muon-hadron pairs. Therefore, for $P+HF$ events,
 the rate of $OS-SS$ pairs is compared to that predicted by the heavy flavor simulation 
 in which pions and kaons are weighted with the corresponding probabilities of mimicking
 a muon signal. In $P+HF$ events,  the number of sequential semileptonic decay-candidates
  ($29262 \pm 850$) is correctly modeled by the rate of sequential decays
 of single $b$-quarks predicted by the simulation ($29190 \pm 1236$).
 This number is 2.5\% of the $P+HF$ total contribution and ($6.9\pm 0.4$)\% of the $b\bar{b}$ contribution
 ($424506 \pm 18454$ events).

 In the remaining $295481 \pm 9271$ ghost events, the number of additional muons in an angular cone
 with $\cos \theta \geq 0.8$ around a primary muon is $49142 \pm 519$.
 In the absence of a simulation of ghost events, that study assumes
  that tracks in ghost events are a 50-50\% mixture of pion and kaons,
 and estimates the number of misidentified additional muons to be $20902 \pm 284$.
 The resulting number of unaccounted ghost events with three or more muons is
  $27970 \pm 538$, ($9.5 \pm 0.4$)\% of the ghost events.

 As shown by Table~\ref{tab:tab_9}, one half of the ghost events
 arise from heavy flavor production acquired with a misidentified muon.
 This type of event should contain an appreciable fraction of additional 
 muons due to semileptonic decays of heavy quarks.
 This contribution was not included in the estimate of Ref.~\cite{a0disc}.

 We estimate this contribution using 
 $K^0_S$ and $K^{*0}$ candidates due to $\pi \rightarrow \mu$ and  $K \rightarrow \mu$ 
 misidentification, respectively.
 As shown in Sec.~\ref{sec:ss-tkaon}, there are $32445 \pm 421$ and
 $87471 \pm 2217$ candidates, respectively.
 We measure the fraction of these candidates in which at least one of the primary muons is accompanied
 by an additional muon in a  $\cos \theta \geq 0.8$ angular cone around its direction.
 We also estimate the contribution of fake additional muon by weighting all hadronic tracks
 that pass the additional muon selection criteria, assumed to be a 50-50\% mixture of pions and kaons,
 with the corresponding misidentification probabilities~\cite{a0disc}.
 
\begin{figure}[htp]
 \begin{center}
 %\vspace{-0.2in}
 \leavevmode
% T1 and F1 in plot_note.root
 %\epsfxsize \textwidth
 \includegraphics*[width=0.5\textwidth]{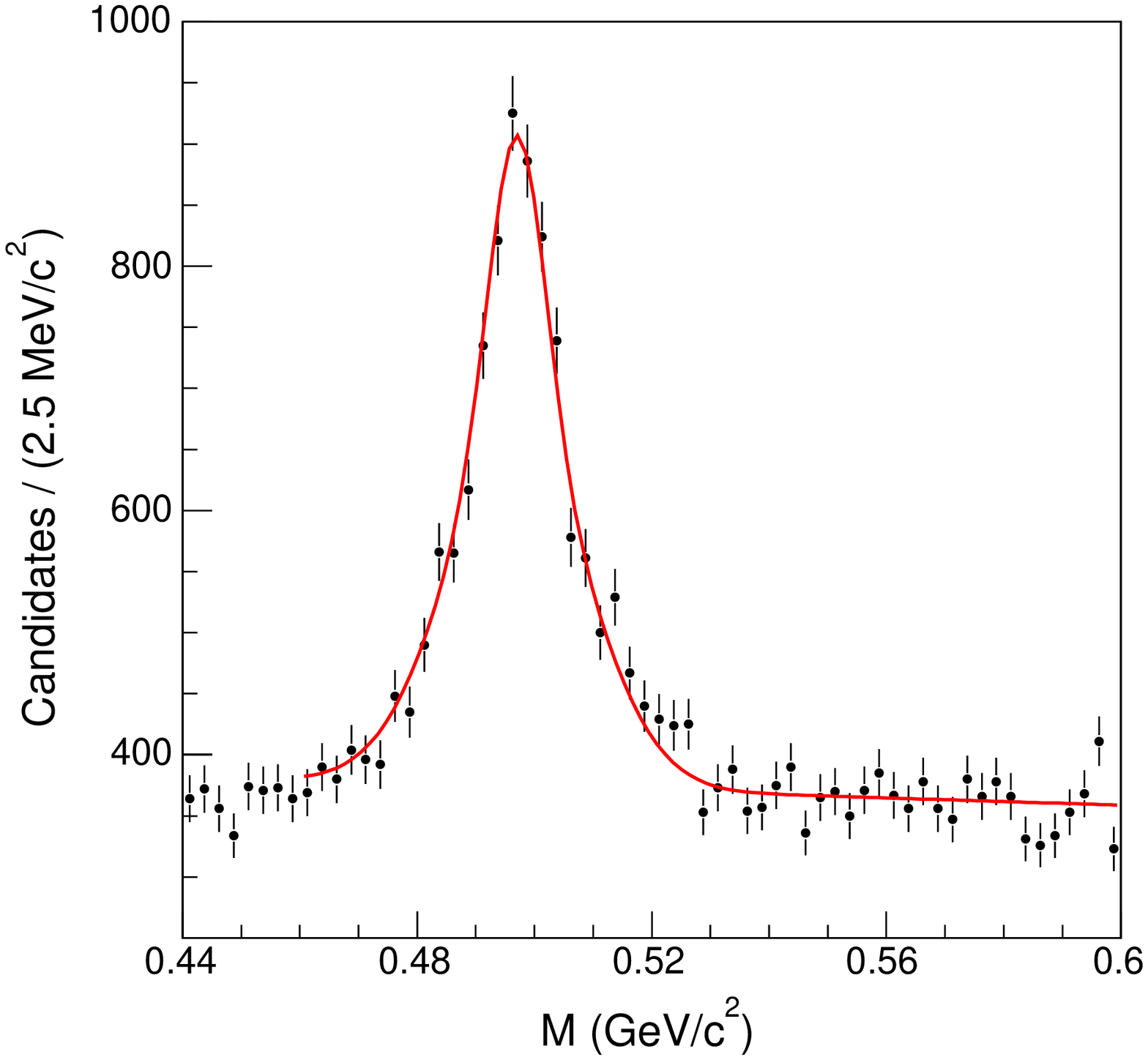}\includegraphics*[width=0.5\textwidth]{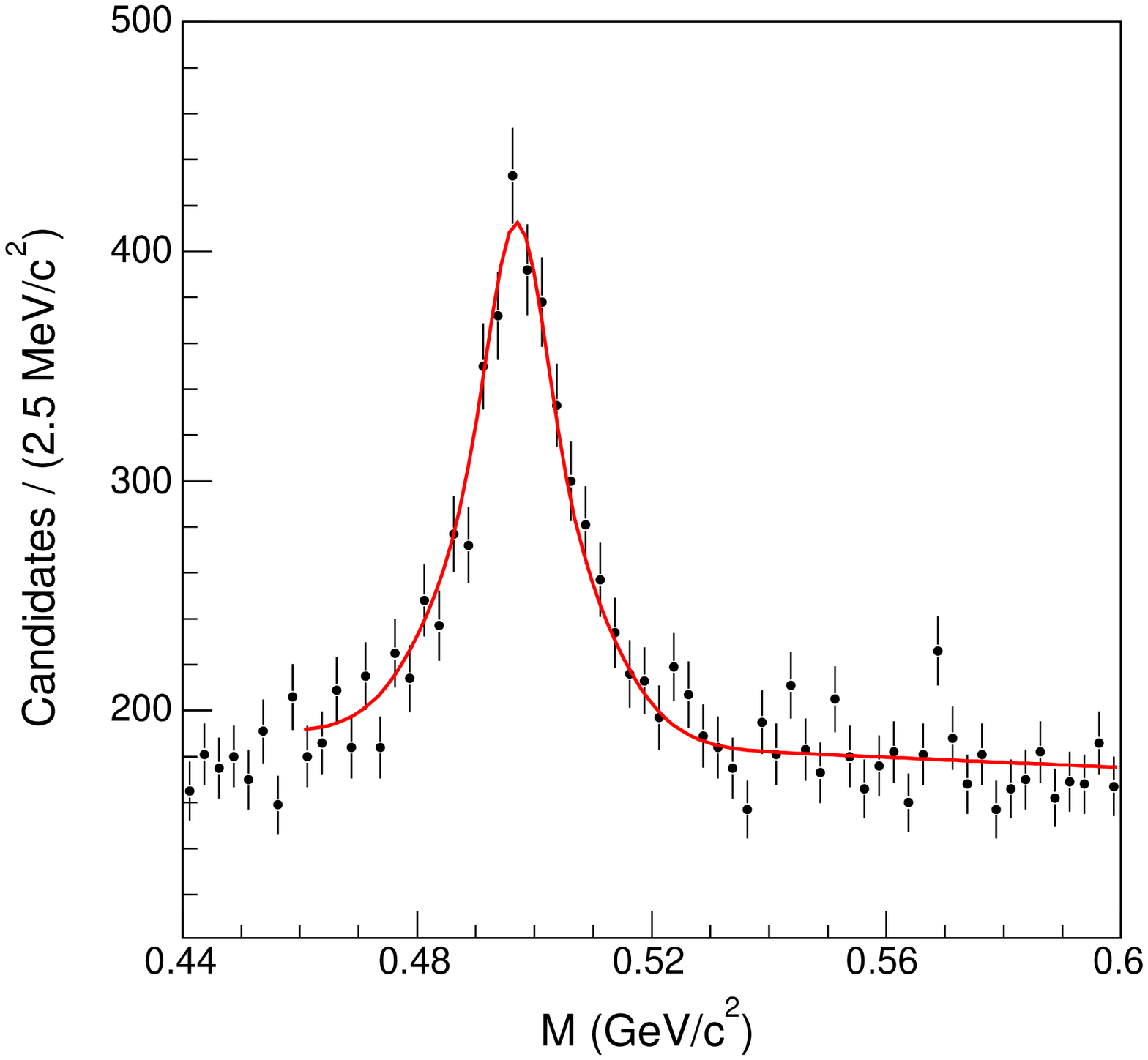}
 \caption[]{Invariant mass distribution of $K^0_S$ candidates
          accompanied by (left) an additional muon and (right) a misidentified muon.
 Lines represent the fits described in the text.}
 \label{fig:fig_6}
 \end{center}
 \end{figure}

\begin{figure}[htp]
 \begin{center}
 \vspace{-0.2in}
 \leavevmode
% T2 and  F2 in plot_note.root
 %\epsfxsize \textwidth
 \includegraphics*[width=0.5\textwidth]{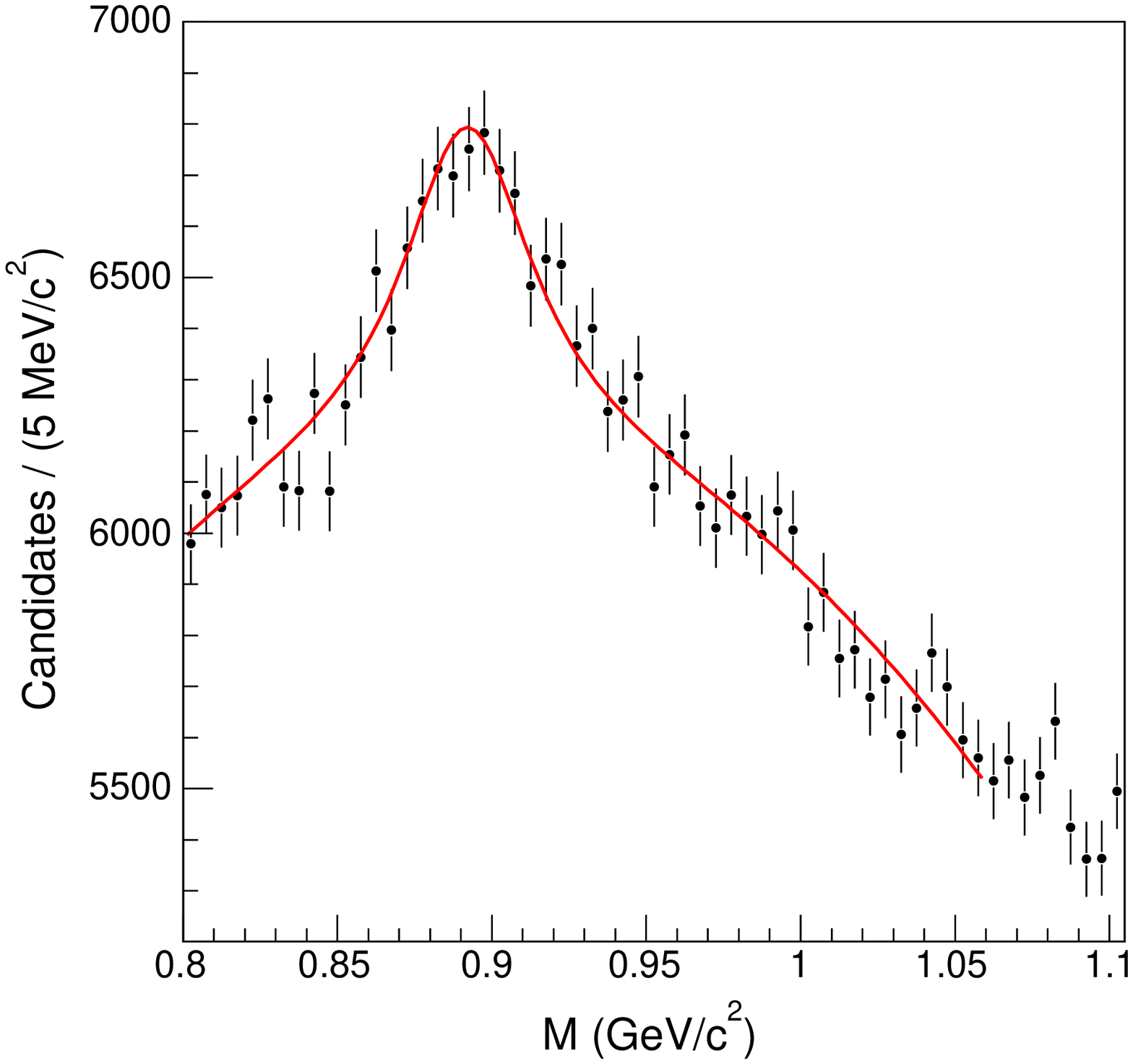}\includegraphics*[width=0.5\textwidth]{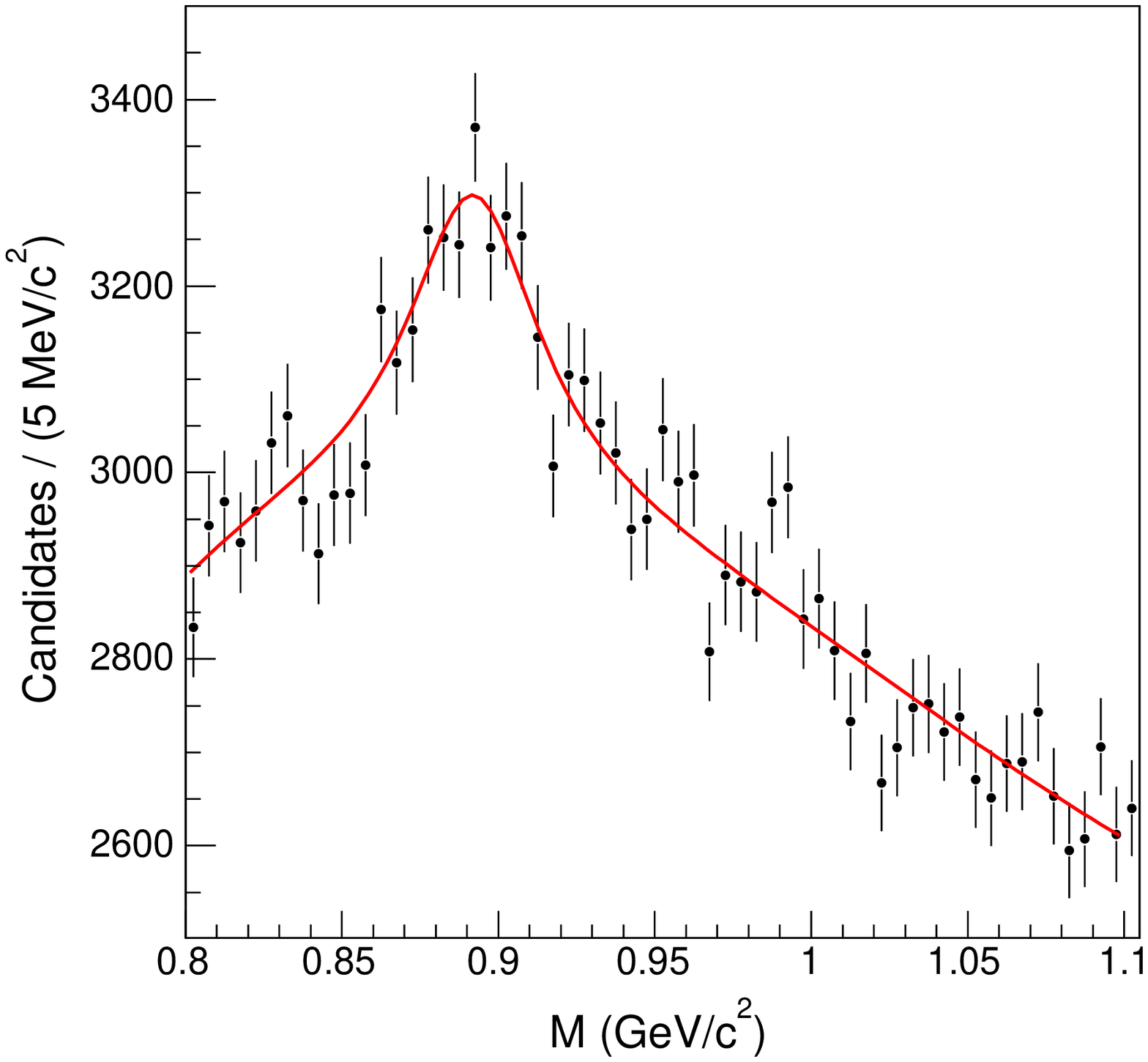}
 \caption[]
{Invariant mass distribution of $K^{*0}$ candidates
          accompanied by (left) an additional muon and (right) a misidentified muon.
 Lines represent the fits described in the text.}
 \label{fig:fig_7}
 \end{center}
 \end{figure}

 Figure~\ref{fig:fig_6} (\ref{fig:fig_7}) shows the invariant mass distribution of 
  $K^0_S$ ($K^{*0}$) candidates when at least one primary muon is accompanied
 by  an  additional  muon or a  predicted misidentified  muon.
 %%%%%%%%%%%%%%%%%%%%%%%%%%
%%%%%%%%%%%%%%%%%%%%%%%%%%%%%%%%%%%%%%%% 
 %%%%%%%%%%%%%%%%%%%%%%%%%%
%%%%%%%%%%%%%%%%%%%%%%%%%%%%%%%%%%%%%%%% 
As previously done, we fit the $K_S^0$ distributions with two Gaussian functions
 to model the signal and a straight line to model the background.
 The  $K^{*0}$ distribution is fitted with a Breit-Wigner function plus a fourth order
 polynomial.

 The fits return $4572 \pm91$ and $1954 \pm 109$ events in which a $K_S^0$ meson 
 is accompanied by an additional muon and by a fake muon, respectively.
 The fits return $10176 \pm 739$ and $5230 \pm 493$ events in which a $K^{*0}$ meson 
 is accompanied by an additional muon and by a fake muon, respectively.

\begin{figure}[htp]
 \begin{center}
 \vspace{-0.2in}
 \leavevmode
% V1 in k0s_puth.root
 %\epsfxsize \textwidth
 \includegraphics*[width=0.5\textwidth]{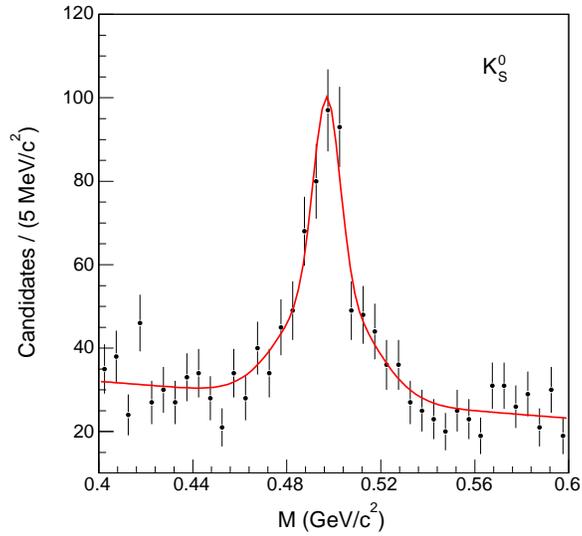}
 \caption[]
{Invariant mass distribution of $K^0_S$ candidates reconstructed using primary and additional muons.
 The line represents the fit described in the text.}
 \label{fig:fig_8}
 \end{center}
 \end{figure}

 As shown in Fig.~\ref{fig:fig_8} 
 for events triggered by  $K_S^0$ misidentifications, sometimes the additional muon is
 contributed by the second prong of the $K_S^0$ decay. 
 Figure~\ref{fig:fig_8} shows the invariant mass distribution
 of primary and additional muons that pass the analysis selection.
 The usual fit yields $403 \pm33$ events in which the additional muon is mimicked by the second leg
 of the $K^0_S$ decay that also produced the primary muon. We remove this contribution
 to evaluate the fraction of real muons accompanying $K \rightarrow \mu$ misidentifications.
 We will add it for the fraction of events triggered by misidentified  $K_S^0$ decays.
 %%%%%%%%%%%%%%%%%%%%%%%%%%
%%%%%%%%%%%%%%%%%%%%%%%%%%%%%%%%%%%%%%%% 

 After removing the predicted numbers of fake muons, the fraction of events  with an identified
  $K_S^0$ meson that contain additional real muons is ($6.83 \pm 0.45$)\%.
 It is ($5.6 \pm 1.05$)\% for events  with an identified
  $K^{*0}$ meson.
  The average of the two fractions is ($6.6 \pm 0.4$)\%. We multiply this fraction
  by the number of ghost events  due to ordinary sources ($241507 \pm 10284$) to
 predict the number of real muons in events due to heavy flavor that are classified as ghost events
 because one of the primary muons was produced by a pion or kaon in-flight-decay.
 This procedure yields a slight overestimate because,
 according to the simulation tuned with the data, ($53.4 \pm 0.4$)\% of the $K \rightarrow \mu$ misidentifications are due to 
 events with heavy flavors, whereas ($51.3 \pm 0.6$)\% of the ghost events arise from heavy flavor production.
 As shown by Table~~\ref{tab:tab_10}, 
  the improved estimate still does not account for  $12169 \pm 1319$ ghost events
 with  additional real muons. 
%%%%%%%%%%%%%%%%%%%%%%%%%%%%%%%%%%
 \begin{table}[htp]
 \caption[]{Number of additional muons in ghost events are compared
  to the  number of expected fake muons and real muons from heavy flavor decays.
 The data correspond to an integrated luminosity of 1426 pb$^{-1}$.}
 \begin{center}
 \begin{ruledtabular}
 \begin{tabular}{cccc}
 %\hline
%\hline 
  Data  &  Fakes &  $K_S^0$ second leg & Heavy Flavor    \\
  $49142 \pm 519$     & $20902 \pm 284$ & $147 \pm 15$ & $15924 \pm 1179$  \\
% \hline
%\hline  
\end{tabular}
 \end{ruledtabular}
 \end{center}
 \label{tab:tab_10}
 \end{table}
%%%
 %%%%%%%%%%%%%%%%%%%%%%%%%%
\section{Conclusions}\label{sec:ss-concl} 
%%%%%%%%%%%%%%%%%%%%%%%%%

 This article reports an improved undestanding of the dimuon samples
 acquired by the CDF experiment.
 One dataset, corresponding to an integrated luminosity of 742 pb$^{-1}$,
 consists of 743006 events
 containing two central ($|\eta|<0.7$) primary (or trigger) muons, each with transverse
 momentum $p_T \geq 3 \; \gevc$, and with invariant mass larger than 
 5 $\gevcc$ and smaller than 80 $\gevcc$.
 These data are split into two subsets: one, referred to as $P+HF$,  consisting of
 $589015\pm 5074$ events in which both muons
 originate inside the beam pipe of radius 1.5 cm; and one, referred to as ghost,
 consisting of $153991 \pm 5074$  events  in which at least one muon
  originates beyond the beam pipe.
 The study in Ref.~\cite{bbxs} shows that the number and properties of $P+HF$ events
 are correctly modeled by the expected contributions of
 semileptonic heavy flavor decays, prompt
 quarkonia decays, Drell-Yan production, and instrumental backgrounds
 due to punchthrough of prompt or heavy-flavored hadrons which mimic a  muon signal.   
 A previous study~\cite{a0disc} has investigated significant
  sources of ghost events, such as in-flight-decays of pions and kaons and
 hyperon decays. That study could account for approximately half of the ghost events
 but was unable to asses the uncertainty of the in-flight-decay prediction.  
 The present study shows that the {\sc herwig} parton-shower generator
 provides an accurate model of the data. The large discrepancy in the previous
 study was generated by not including the contribution of final states in which
 a  $b(c)$ hadron decays semileptonically and the second muon is produced by
 the in-flight-decay of a particle in the recoiling jet.
 After tuning by a few percent the pion and kaon rates predicted by the simulation 
 with a fit to the data, we  show that ordinary sources, mostly in-flight-decays,
 account for $125665 \pm 5351$  of the
 $153991 \pm 5074$ ghost events isolated in the sample of 743006 dimuons.

 For comparison, a D0 study has used a similar dimuon sample to set a limit~\cite{williams} 
 of $(0.4\pm0.26\; {\rm stat}\pm0.53\; {\rm syst})$\% to the fraction of muons
 produced at a distance larger than 1.6 cm from the beamline including pion and kaon
 in-flight-decays.
 This appears to be in contradiction with the present result, and also with a recent
 estimate~\cite{d0cp} of the fraction  of 
 $K \rightarrow \mu$ and $\pi \rightarrow \mu$
 contributions in the D0 subset of same-charge dimuons ($\simeq 40$\%). 

 The present study  also  improves a previous estimate~\cite{a0disc}
 of the content of additional muons
 with $p_T \geq 2\; \gevc$ and $|\eta| \leq 1.1$ in ghost events.
 We find that ($23 \pm 6$)\% of the unaccounted ghost events contain additional
 real muons. For comparison, the fraction of  $b\bar{b}$ events that
 contain  additional muons due to sequential semileptonic
 decays is ($6.9 \pm 0.4$)\%. 

 Both results presented in this article have implications for measurements 
 derived in dimuon datasets 
 without properly accounting for the presence of ghost events. As an example,  
 the measurement of the dimuon charge asymmetry performed by the D0 experiment~\cite{d0cp}
 estimates the fraction of $K \rightarrow \mu$ and $\pi \rightarrow \mu$ misidentifications
 with a similar method.
 After removing this background,
  the remaining muon pairs with same charge
 are attributed to $b\bar{b}$ production. The present study shows that, after removing
 this type of misidentified muons,
 the  data set still contains an additional component that cannot be accounted for
 with ordinary sources. The size of this component, equally
 split in opposite and same sign pairs~\cite{a0disc}, is
 ($12.8 \pm 3.2$)\% of the total number of dimuons due to $b\bar{b}$ production.
%%%%%%%%%%%%%%%%%%%%%%%%%%%%%%%%%%%%%%%%%%%%
 \section{Acknowledgments}
%%%%%%%%%%%%%%%%%%%%%%%%%%%%%%%%%%%%%%%%%%%

 We thank the Fermilab staff and the technical staffs of the participating institutions for their vital contributions.
 This work was supported by the U.S. Department of Energy and National Science Foundation;
 the Italian Istituto Nazionale di Fisica Nucleare; the Ministry of Education,
 Culture, Sports, Science and Technology of Japan;
 the Natural Sciences and Engineering Research Council of Canada; the National Science Council of the Republic of China;
 the Swiss National Science Foundation; the A.P. Sloan Foundation;
 the Korean World Class University Program, the National Research Foundation of Korea; 
 the Science and Technology Facilities Council and the Royal Society, UK;
 the Institut National de Physique Nucleaire et Physique des Particules/CNRS; the Russian Foundation for Basic Research;
 the Ministerio de Ciencia e Innovaci\'{o}n, and Programa Consolider-Ingenio 2010, Spain;
 the Slovak R\&D Agency; the Academy of Finland; and the Australian Research Council (ARC). 
%%%%%%%%%%%%%%%%%%%%%%%%%%%%%
%%%%%%%%%%%%%%%%%%%%%%%%%%%%%%%%%%%%%%%%%%%%%%%%%%%%  

%%%%%%%%%%%%%%%%%%%%%%%%%%%%%%%
%%%%%%%%%%%%%%%%%%%%%%%%%%%%%%%

\begin{thebibliography}{99}
\bibitem{d0}      The impact parameter is defined as the distance of closest
                  approach of a track to the primary event vertex in the 
                  transverse plane with respect to the beamline.
\bibitem{strange} That study follows the methodology of previous measurements
                  that ignored other possible sources of muons. For example,
                  muon tracks from pion and kaon in-flight-decays inside the
                  tracking volume were regarded as prompt tracks because the
                  track reconstruction algorithms were believed to remove
                  decay muons with an appreciable kink.
\end{thebibliography}
 \end{document}